\documentclass[12pt,twoside]{article}
\usepackage{enumerate}
\usepackage{amsthm,amsmath,amssymb,amscd} %latexsym,
\usepackage{amsmath,amsthm,amssymb,graphicx}
\usepackage{graphicx}
\usepackage{subfigure}
\newtheorem{theorem}{Theorem}

\newtheorem{proposition}{Proposition}
\theoremstyle{definition}

\def \beq{\begin{equation}}
\def \eeq{\end{equation}}
\begin{tiny}\end{tiny}

\def\.#1{\dot #1}

\def \( {\big( }
\def \) {\big) }

\def \bar{\overline} 
\def \1{{\bf I}}
\def \2{{\bf II}}
\def\.#1{\dot #1}

\begin{document}
\DeclareGraphicsRule{.png}{eps}{.bb}{`bmeps #1}
\title{Linear stability of the Lagrangian triangle  solutions for quasihomogeneous potentials.}

\author{Manuele Santoprete\thanks{Department of Mathematics, University
of California, Irvine, 294 Multipurpose Science \& Technology
Building, Irvine CA, 92697 USA.
E-mail: msantopr@math.uci.edu}
}

\maketitle

\begin{abstract}
\noindent In this paper we study the linear  stability of the  relative equilibria  for homogeneous and quasihomogeneous potentials. Firstly, in the case the potential is a homogeneous function of degree $-a$, we find that any relative equilibrium of the $n$-body problem with $a>2$ is spectrally unstable. We also find a similar condition in the quasihomogeneous case.
Then we consider the case of three bodies and we study the stability of the equilateral triangle relative equilibria. In the case of homogeneous potentials we recover the classical result obtained by Routh in a simpler way. In the case of quasihomogeneous potentials we find a generalization of Routh inequality and we show that, for certain values of the masses, the stability of the relative equilibria depends  on the size of the configuration.  
\end{abstract}
%\markboth{M. Santoprete}{Stability}

%\pacs{}
%\ams{70F15, 70F10, 70F07}
 
homogeneous potentials, quasihomogeneous potentials.

%%%%%%%%%%%%%%%%%%%%%%%%%%
\section{Introduction}
%%%%%%%%%%%%%%%%%%%%%%%%%%
The study of the stability of the relative equilibria is very important in the analysis of Hamiltonian systems. In the last few decades some general methods have been developed to tackle this problem, namely the {\it energy-Casimir method} and the {\it energy momentum method}. The energy-Casimir method is an adaptation of the classical Lagrange-Dirichlet method \cite{Marsden} that was developed by  Arnold \cite{Arnold66}
and applied to analyze the stability of stationary flows of perfect fluids. Such method was also applied to a variety of other problems. However this technique relies heavily on a supply of  Casimir functions. The energy-momentum method was developed to overcome this difficulty, it is  very powerful and was applied successfully to many problems.

Unfortunately there are several problems that cannot be tackled using the energy momentum method. Indeed in some cases the second variation of the energy  is indeterminate. In particular it is known that in celestial mechanics the stability of the relative equilibria cannot be studied using the energy momentum method. However one can study the linear or spectral stability.  The actual stability of relative equilibria is a major open problem in celestial mechanics. Several authors  (see for example \cite{Moeckel94}) think that all the relative equilibria are nonlinearly unstable because Arnold's diffusion is believed to be a feature of the $n$-body problem. Remarkably it is very difficult to study Arnold's diffusion numerically and the existence of Arnold's diffusion in the $n$-body problem has not been proved beyond doubt.

In this paper we consider the problem of $n$ bodies interacting by means of  homogeneous and quasihomogeneous potentials. In the case the potential is a homogeneous function of degree $-a$ we find that any relative equilibrium is unstable if $a>2$. In the case of quasihomogeneous potentials we find a similar, but more complex condition. 
Then we study the equilateral triangle relative equilibria and  their stability. The energy-momentum method fails to provide  informations concerning the stability of the relative equilibria since the second variation of the amended potential is indefinite. Therefore we study  the linear stability of those solutions. 
The linear stability of the Lagrangian  triangle  solution of the Newtonian three body problem was first studied by Gascheau in 1843 \cite{Gascheau}.
Later  Routh \cite{Routh} studied the linear stability of the same solutions in the case of homogeneous potentials.
He proved that linear stability is achieved only when $a<2$ and  the masses satisfy
\beq
\frac{m_1m_2+m_1m_3+m_2m_3}{(m_1+m_2+m_3)^2}<\frac{1}{3}\left (\frac{2-a}{2+a}\right)^2
\label{routhinequality0}
\eeq
where the potential is a homogeneous function of degree $-a$ with $a>0$.
In this work, using a considerably simpler derivation based on an idea of Moeckel (see \cite{Moeckel95}),  we recover the classical result of Gascheau and Routh (for spectral stability),  and we analyze it. In particular we find  that the  equilateral triangle relative equilibria can be spectrally stable for $0<a<2$ if one of the masses is much larger than the other two. An interesting feature  appears when $0<a<14-8\sqrt{3}$, indeed, in this interval, there is spectral stability not only in presence of a dominant mass but also when one of the masses is sufficiently small. This is a somewhat surprising result that stresses the difference between the Newtonian potential and other homogeneous potentials. Indeed it is widely believed that, in the Newtonian case, as Moeckel conjectured \cite{Moeckel94}, a relative equilibrium can be linearly stable only if it contains a mass significantly larger than the other masses.

Then we use similar methods to generalize Routh condition to the case of quasihomogeneous potentials. Quasihomogeneous potentials (\cite{Diacu96,Perez98}) have been widely studied in recent years since they are suitable to describe a number of phenomena. Among the other we recall some of the most representative:  the so called Manev (see \cite{Diacu93,Diacu95,Delgado,Llibre} and references therein)  and Schwarzschild (see \cite{Schwarzschild,Wald}) potentials. The Manev potential and the Schwarzschild  potentials are particularly interesting because, under suitable conditions, they  describe the motion of a test particle in a gravitational field. In particular it was shown that the Manev potential can describe the precession of the perihelion of Mercury with the same accuracy as general relativity (see \cite{Hagihara}). The Schwarzschild potential can also be used for similar calculations (for example see \cite{Wald}). %In particular  the Schwarzschild potential is often used in General Relativity (for example see \cite{Wald}).

The Lennard-Jones potential (\cite{Corbera04}) is also a very interesting potential   used in molecular dynamics to simulate many particle systems as for example solids, liquids  and gases. The relative equilibria for this potential  were  studied in \cite{Corbera04} and  will not be considered here.

In the case of quasihomogeneous potentials we find that spectral stability is achieved only when 
\[
\frac{(a^2-2a)r_0^{b+2}+(b^2-2b)r_0^{a+2}}{ar_0^{b+2}+br_0^{a+2}}<0
\]
and 
\beq
\frac{m_1m_2+m_1m_3+m_2m_3}{(m_1+m_2+m_3)^2}\leq\frac{1}{3}\left (\frac{b(b-2)r_0^{a-b}+a(a-2)}{b(b+2)r_0^{a-b}+a(a+2)}\right)^2
\label{generalrouthinequality0}
\eeq
where $r_0$ is the mutual distance between any two  bodies and the potential is the sum of a homogeneous function of degree $-a$ and of one of degree $-b$ with $a>b>0$. We denote with $f$ the left hand side of inequality (\ref{generalrouthinequality0}).

Analyzing the conditions above we find that, for certain values of the masses, the stability of the Lagrangian triangle solutions depend on the size of the triangle. In particular we show that if $a>b>2$ the triangle solutions are unstable for any value of the masses. If $a>2>b>0$ the above solutions are unstable when $f>1/3[(b-2)/(b+2)]^2$ and when 
$f\leq 1/3[(b-2)/(b+2)]^2$ and $r_0<z_1^*$ (for some $z_1^*\in\mathbb{R}^+$). Moreover they are spectrally stable when $f\leq 1/3[(b-2)/(b+2)]^2$ and $r_0\geq z_1^*$.  If $2>a>b>0$ there are three different cases depending on the value of the masses.
When $f>1/3[(b-2)/(b+2)]^2$ the solutions are unstable, while when $f\leq 1/3[(a-2)/(a+2)]^2$ they are spectrally stable. When $1/3[(a-2)/(a+2)]^2<f\leq 1/3[(b-2)/(b+2)]^2$ they are unstable for $r_0<r_0*$, but spectrally stable for $r_0\geq r_0^*$ for some $r_0^*\in\mathbb{R}^+$.

The study of the stability of relative equilibria for quasihomogeneous potentials is quite interesting. Indeed the Trojan asteroids are found at the Lagrangian point of the Sun-Jupiter system. The gravitational interaction is usually considered to be the Newtonian one in most of the studies concerning the Trojans. However if one wants to consider the problem in the framework of General  Relativity (see \cite{Rosswog} for a study of the three-body problem in the Post-Newtonian approximation of General Relativity ) then one might get some surprising results. Since the Schwarzschild and the Manev potentials are used to approximate general relativity the results presented in this paper suggest that it might be impossible to find, in nature, equilateral triangle relative equilibria smaller than a certain size. This is because the relative equilibria  studied in this work (in the case of quasihomogeneous potentials) are unstable if the interacting bodies are too close to each others.

This paper is organized as follows. In the next section we write the equation of motion in Cartesian coordinates and  in uniformly rotating coordinates. We define the relative equilibria for the $n$-body problem and we formulate the conditions for their spectral stability.  In section 3 we study the case of homogeneous potentials. First we consider the $n$-body problem in general  and then for $n=3$ we recover the inequality of Routh to describe the stability of the Lagrange triangle relative equilibria. 
In the last  section we consider quasihomogeneous potentials. We first study the $n$-body problem  and then  we find a condition for the spectral stability of the Lagrange triangle solutions and we discuss it. 
%%%%%%%%%%%%%%%%%%%%%%%%%%%%%%%%%%
\section{Relative Equilibria and Their Stability}
%%%%%%%%%%%%%%%%%%%%%%%%%%%%%%%%%%%
%%%%%%%%%%%%%%%%%%%%%%%%%%%%%%%%%%
\subsection{Relative Equilibria}
%%%%%%%%%%%%%%%%%%%%%%%%%%%%%%%%%%%
We let the mass and position of the $n$ bodies be given by $m_i\in {\mathbb R}^+$ and ${\bf q}_i\in {\mathbb R}^2$, where $i=1,\ldots,n$. Let $q=({\bf q}_1,\ldots,{\bf q}_n)\in{\mathbb R}^{2n}$. Newton's equations for the $i$th body are
\[
m_i\ddot{\bf q}_i=\frac{\partial {\cal U}}{\partial {\bf q}_i}
\] 
where ${\cal U}(q)$ is the potential function.
We let the momentum of each body be ${\bf p}_i=m_i\dot{\bf q}_i$ and let $p=({\bf p}_1,\ldots,{\bf p}_n)\in{\mathbb R}^{2n}$.
The equations of motion can be written as 
\beq
\begin{split}
\dot q &=M^{-1}p=\frac{\partial H}{\partial p}\\
\dot p &=\nabla {\cal U}(q)=-\frac{\partial H}{\partial q}
\end{split}
\label{eqmotion}
\eeq
where  $M$ is the $2n\times 2n$  mass matrix $\mbox{diag}(m_1,m_1,\ldots,m_n,m_n)$, $\nabla$ denotes the Euclidean gradient in ${\mathbb R}^{2n}$ and $H(q,p)$ is the Hamiltonian function
\[
H(q,p)=\sum_{i=1}^n\frac{\|{\bf p}_i\|^2}{2m_i}-{\cal U}(q)=\frac 1 2 p^TM^{-1}p -{\cal U}(q).
\]
Let $J$ and $R(\theta)$ denote the $2n\times2n$ block diagonal matrices with $n$ equal $2\times 2$ blocks of the form
\[ K=\left [
\begin{array}{cc}
0&1\\
-1&0
\end{array}\right ]\quad \mbox{and} \quad e^{K\theta}=\left [
\begin{array}{cc}
\cos\theta&\sin\theta\\
-\sin\theta&\cos\theta
\end{array}\right ]
\]
respectively.
To introduce coordinates that uniformly rotate with constant angular velocity $\hat\omega$, we let ${\bf x}_i={\bf q}_ie^{\hat\omega K t}$ and ${\bf y}_i={\bf p}_ie^{\hat\omega K t}$. This is a symplectic change of variable and thus preserves the Hamiltonian structure of system (\ref{eqmotion}).
The equations of motion in the new variables are
\beq
\begin{split}
\dot x &=\hat\omega J x+M^{-1}y=\frac{\partial  \hat H}{\partial y}\\
\dot y &= \nabla{\cal U}(x)+\hat\omega J y=-\frac{\partial \hat H}{\partial x}
\end{split}
\label{eqmotionrot}
\eeq
where $x(t)=R(\hat\omega t)q(t)$, $y(t)=R(\hat\omega t)p(t)$ and $\hat H(x,y)$ is the Hamiltonian function:
\[
\hat H(x,y)=\frac 1 2 y^TM^{-1}y-{\cal U}(x)-\hat\omega x^TJy.
\]
An equilibrium point of system (\ref{eqmotionrot}) corresponds to a periodic solution in the $n$-body problem consisting of a configuration of masses which rotates rigidly about its center of mass. 
An equilibrium $(x,y)$ of system (\ref{eqmotionrot}) must satisfy $y=-\hat\omega M J x$ and
\beq
\nabla{\cal U}(x)+\hat\omega^2Mx=0
\label{releq}
\eeq
A {\it relative equilibrium} is a configuration $x\in{\mathbb R}^{2n}$ which satisfies the algebraic equation (\ref{releq}) for some value of $\hat\omega$.
This can be viewed as the condition for a critical point of the restriction of the potential to $\{x\in{\mathbb R}^{2n}:x^TMx=c\}$ for any constant $c$, where $\hat\omega^2$ plays the role of a Lagrange multiplier.

In the following we will consider the potential ${\cal U}$ to be either a homogeneous potential of the form
\beq
V(q)=\sum_{i<j}\frac{m_im_j}{\|{\bf q}_i-{\bf q}_j\|^a}
\label{homa}
\eeq
and
\beq
W(q)=\sum_{i<j}\frac{m_im_j}{\|{\bf q}_i-{\bf q}_j\|^b},
\label{homb}
\eeq
where $0<b<a$, or a quasihomogeneous potential defined as a sum of the homogeneous potentials above:
\beq
U(q)=V(q)+W(q).
\label{quasihom}
\eeq
A configuration $x$ is a relative equilibrium for the quasihomogeneous potential $U$ provided that
\beq
\nabla U(x)=-\omega^2Mx
\eeq
for some constant $\omega$, while it is a relative equilibrium for $V$ and $W$ if
\beq
\nabla V(x)=-\omega_1^2Mx \quad \mbox{and}\quad \nabla W(x)=-\omega_2^2Mx
\label{hom}
\eeq
respectively.
Since the potentials $V$ and $W$ are homogeneous of degree $-a$ and $-b$ respectively the   constants $\omega_1$ and $\omega_2$ are determined by: 
\beq
\omega_1^2=\frac{aV(x)}{x^TMx} \qquad \omega_2^2=b\frac{W(x)}{x^TMx}
\eeq
Furthermore if $x$ is simultaneously a relative equilibrium for $V$, $W$ and $U$ then 
\[\nabla U(x)=\nabla V(x)+\nabla W(x)=-(\omega_1^2+\omega_2^2)Mx \]
and 
\[
\omega^2=\omega_1^2+\omega_2^2.
\]
%%%%%%%%%%%%%%%%%%%%%%%%%%%%%%%%%%
\subsection{Linear and Spectral Stability }
%%%%%%%%%%%%%%%%%%%%%%%%%%%%%%%%%%%
A relative equilibrium $x$ is {\it linearly stable} if the origin is a stable solution of the linearization
at $x$ of system (\ref{eqmotionrot}). A necessary condition for $x$ to be linearly stable is that all the eigenvalues of the linearization
\beq S=\left [
\begin{array}{cc}
\hat\omega J&M^{-1}\\
D\nabla {\cal U}&\hat\omega J
\end{array}\right ]
\eeq
are either zero or purely imaginary. This weaker condition is called {\it spectral stability}.

The characteristic polynomial of $S$, $P(\lambda)$, is of degree $4n$ and  is an even polynomial, since $S$ is a Hamiltonian matrix. Let $v$ be an eigenvector of $S$ with eigenvalue $\lambda$ and write $v=(v_1,v_2)$, where $v_1,v_2\in {\mathbb C}^{2n}$. The eigenvector equation reduces to
\beq
\begin{split}
v_2 &=M(\lambda I-\hat\omega J)v_1\\
{\cal A} v_1 & =0
\end{split}
\eeq
where
\beq
{\cal A}=M^{-1}D\nabla {\cal U}+(\hat\omega^2-\lambda^2)I+2\lambda\hat\omega J
\eeq
and $I$ is the identity matrix.
Consequently, to obtain the eigenvalues of $S$, one needs only take the determinant of ${\cal A}$ and find the roots, namely $P(\lambda)=\det ({\cal A})$.
Following Moeckel we introduce the normalized eigenvalues $\mu=\lambda/|\hat\omega|$. These satisfy the equation
\beq
\det(M^{-1}D\nabla {\cal U}+(1-\mu^2)I+2\mu J)=0.
\label{det}
\eeq
Note that the stability condition is unchanged since the normalization factor is real and positive.

Two vectors $v$ and $w$ are called $M$-orthogonal if $v^TMW=0$. One can show that $J$ is antisymmetric and $M^{-1}D\nabla {\cal U}$ is symmetric in any $M$-orthonormal basis. Using an $M$-orthogonal basis and taking the transpose of ${\cal A}$ does not change the determinant (\ref{det}) and thus the determinant is an even function of $\mu$. Let $z=\mu^2$ and let $G(z)$ be the polynomial of degree $2n$ 
\beq
G(\mu^2)=\det(M^{-1}D\nabla {\cal U}+(1-\mu^2)I+2\mu J)=0.
\label{poly}
\eeq
Then $x$ is spectrally stable if and only if all the roots of $G(z)$ are either zero or real and negative.
$G(z)$ is called the {\it stability polynomial}.

In order to study spectral stability it is convenient to obtain factorizations of the stability polynomial $G(z)$. Moeckel's idea is that such a factorization can be obtained by finding subspaces of ${\mathbb R}^{2n}$ which are simultaneously invariant for $J$ and $M^{-1}D\nabla{\cal U}$:
\begin{proposition}
Suppose that $W\in {\mathbb R}^{2n}$ is a invariant subspace for both   $J$ and $M^{-1}D\nabla{\cal U}$. Then the stability polynomial can be factored into two even polynomials in $G(z)=G_1(z)G_2(z)$ where $G_1(z)$ and $G_2(z)$ are given by (\ref{poly}) with the matrices involved restricted to the subspaces $W$ and $W^\perp=\{v\in{\mathbb R}^{2n}:v^TM w=0~ \forall~ w\in W \}$.
\end{proposition}

The invariant subspaces $W$ must have even dimension. The simplest case is dimension two. In this paper we will only need to use invariant subspaces of dimension two.
 In this case we have the following
\begin{proposition}
Suppose $W\in{\mathbb R}^{2n}$ is a two-dimensional subspace which is simultaneously invariant for $J$ and $M^{-1}D\nabla{\cal U}$. Let the eigenvalues of the restriction of $\hat\omega^{-2}M^{-1}D\nabla{\cal U}$
be $\eta$ and $\xi$. Then $G(z)$ has a quadratic factor
\[
Q(z)=z^2+\alpha z+\beta
\]
where $\alpha=2-\eta-\xi$ and $\beta=(1+\eta)(1+\xi)$. The corresponding roots are real and negative if and only if
\[
\alpha>0\quad \beta>0\quad \alpha^2-4\beta\geq 0. 
\]
\label{prop}
\end{proposition}
%%%%%%%%%%%%%%%%%%%%%%%%%%
\section{Homogeneous Potentials}
%%%%%%%%%%%%%%%%%%%%%%%%%%
\subsection{General Case}
%%%%%%%%%%%%%%%%%%%%%%%%%%%%%%%%%%%%%%
Consider the potential $V(x)$. Then the matrix $D\nabla V(x)$ is of the form:
\[A=\left [\begin{array}{ccc}
A_{11}& \cdots&A_{1n}\\
\vdots   &  &\vdots\\
A_{n1}&\cdots&A_{nn}\\
\end{array}\right ]\]
where $A_{ij}$ is the $2\times 2$ matrix given by
\beq
\begin{split}
A_{jk}&=a\frac{m_im_k}{d_{jk}^{a+2}} \left [I-(a+2) {\bf u}_{jk} {\bf u}_{jk}^T \right] \quad \mbox{if} \quad j\neq k\\
A_{kk}&=-\sum_{j\neq k}A_{jk}
\end{split}
\eeq
where ${\bf u}_{jk}=\frac{{\bf x}_k-{\bf x}_j}{d_{jk}}$ and $d_{jk}=\|{\bf x}_j-{\bf x}_k\|$. 

Using the fact that the diagonal blocks of $A$ are the negative of the sum of the blocks in the corresponding rows it is clear that both $v=(1,0,\ldots,1,0)$ and $w=(0,1,\ldots,0,1)$ are in the kernel of $A$. Therefore  $\mbox{span}\{v,w\}$  is a two-dimensional invariant subspace for both $M^{-1}A$ and $J$.
The eigenvalues of the restriction of $\omega_1^2M^{-1}A$ are $\eta=\xi=0$. Applying Proposition \ref{prop} one finds that the the roots of $Q(z)$ are real and negative. Substituting $\eta=\xi=0$ into $Q(z)$ yields the repeated root $z=-1$. The corresponding eigenvalues are $\mu=\pm i$. These values are a result of  a drift in the center of mass, and are present in any relative equilibrium.

Another two-dimensional invariant subspace comes from the configuration itself. 
Consider $v=x$ and $w=Jx$. Equation (\ref{hom}) with the homogeneity of the potential gives:
\beq
\begin{split}
 M^{-1}D\nabla V(x)v &=-(a+1)M^{-1}\nabla V(x)\\&=(a+1)M^{-1}\omega_1^2Mx=(a+1)\omega_1^2v,
\end{split}
\eeq
so $v$ is an eigenvector of $M^{-1}B$ with eigenvalue $(a+1)\omega_1^2$.
On the other hand $w=(R^{-1}(0))'x$ where $R(\theta)$ is the $2n$-dimensional rotation operator introduced above.
Moreover
\[
\nabla V(R^{-1}(\theta)x)=R^{-1}(\theta)\nabla V(x)
\]
because of the rotation invariance of the potential. Differentiating at $\theta=0$ gives:
\beq
M^{-1}D\nabla V(x)w=M^{-1}J\nabla V(x)=-M^{-1}J\omega_1^2Mx=-\omega_1^2w
\eeq
and thus $w$ is an eigenvector with eigenvalue $-\omega_1^2$. Choosing  $\eta=(a+1)$ and $\xi=-1$ we find that $\alpha=2-a$, and the roots of $Q(z)$ are positive if $a>2$. This proves the following

\begin{theorem}
Any relative equilibrium of $V(x)$  is spectrally unstable if $a> 2$. 
\end{theorem}
%%%%%%%%%%%%%%%%%%%%%%%%%%%%%%%
\subsection{Lagrangian Triangle Relative Equilibria}
%%%%%%%%%%%%%%%%%%%%%%%%%%%%%%%%%
%-------------------------------------------------------------------------------------
\begin{figure}[t]
  \begin{center}
    \mbox{
      \subfigure[]{\resizebox{!}{5.5cm}{\includegraphics{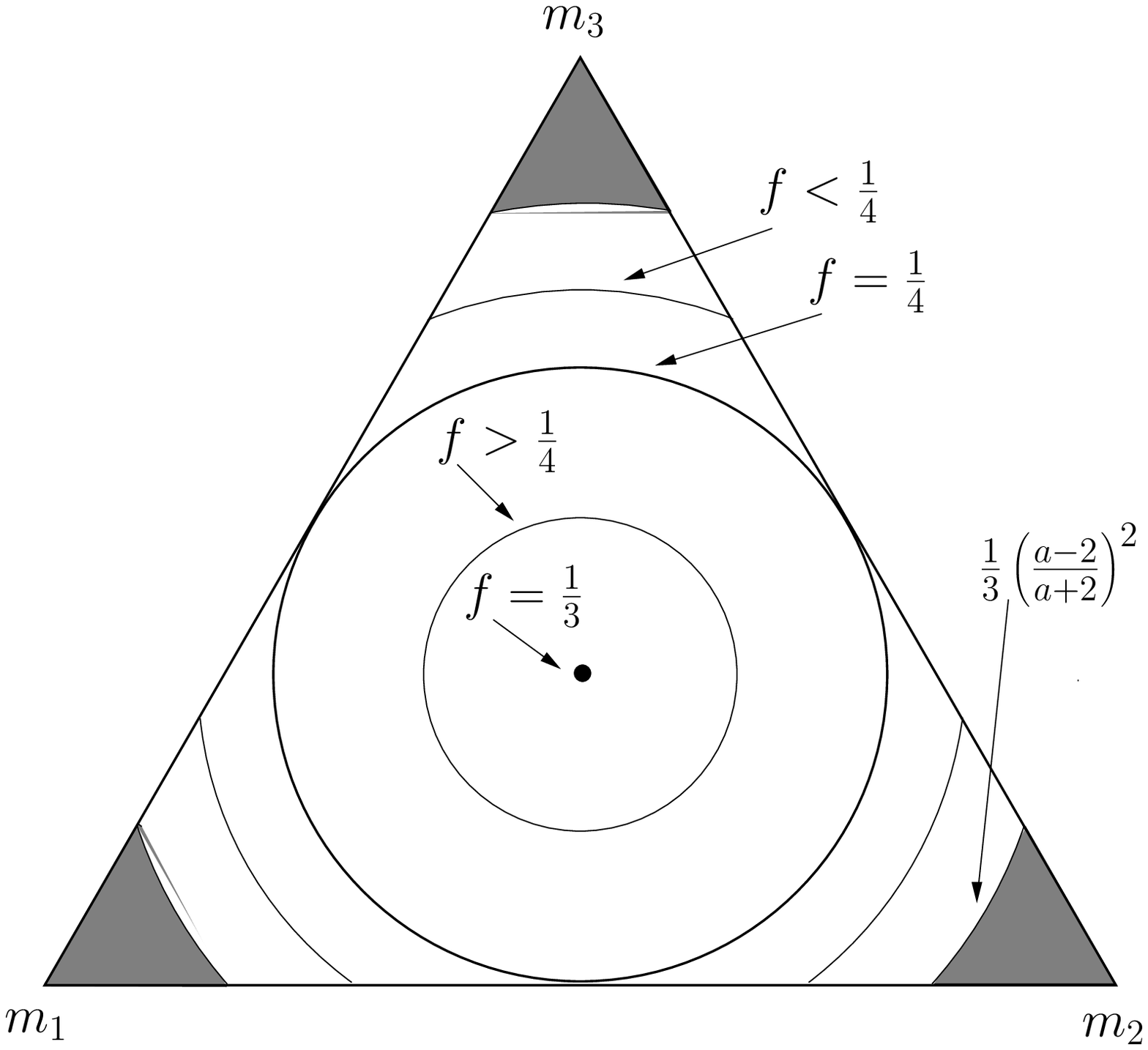}} \label{triangle}} \quad
      \subfigure[]{\resizebox{!}{5.5cm}{\includegraphics{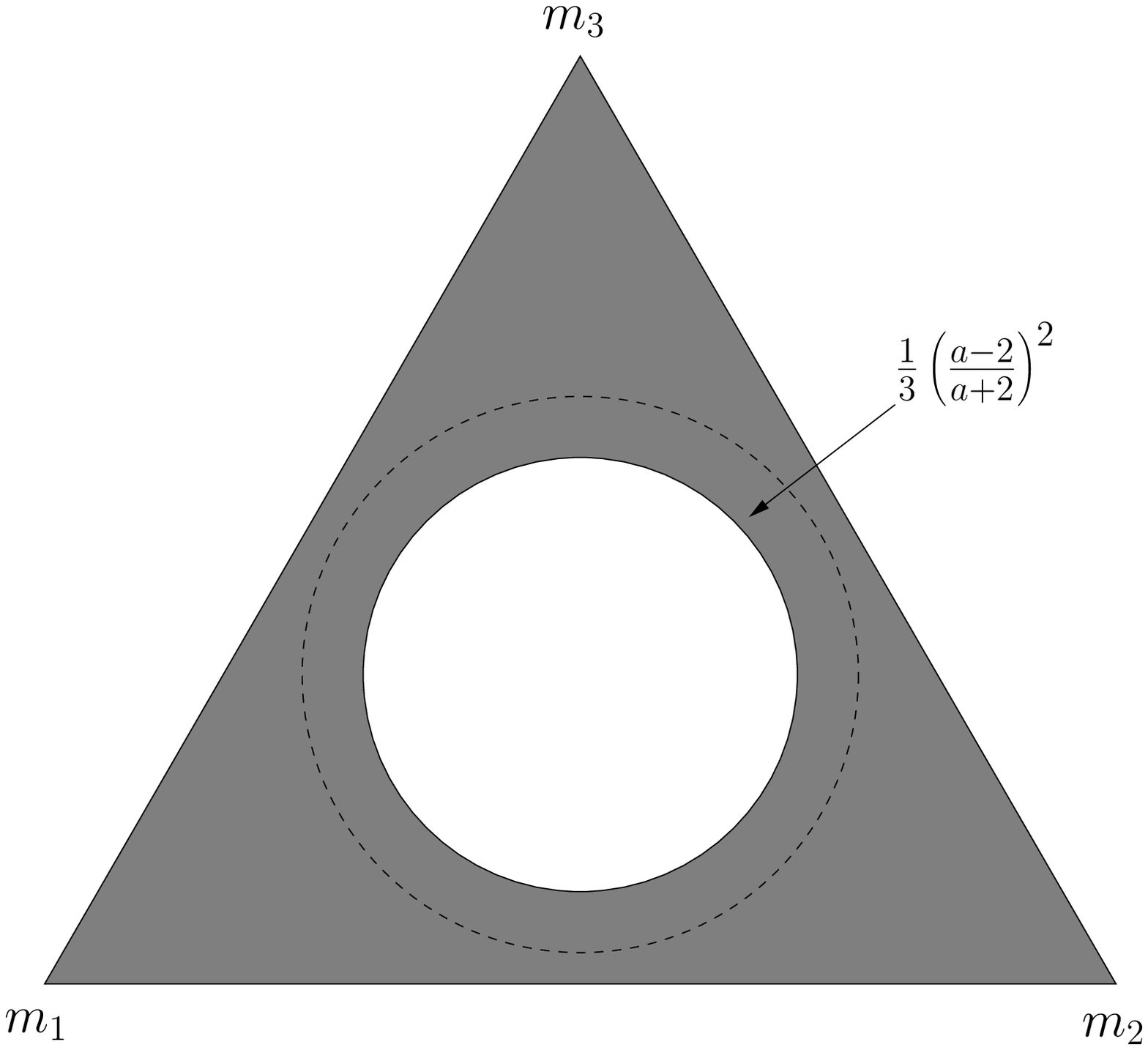}} \label{triangle1}}} 
    \caption{(a) Some level curves $f(m_1,m_2,m_3)=c$ in the triangle of normalized masses. Regions of stability (shaded) and instability in the triangle of normalized masses for the homogeneous potential $V$ with $14-8\sqrt{3}<a<2$. (b) Regions of stability (shaded) and instability  for the  potential $V$ with $0<a<14-8\sqrt{3}$.}
  \end{center}
\end{figure}
%-----------------------------------------------------------------------------------------------

Now let $n=3$. Then the equilateral triangle configurations with the center of mass at the origin are relative equilibria, for any choice of the masses $m_1$, $m_2$ and $m_3$. 
Moreover, since the potential is homogeneous, the stability analysis is not affected by the size of the configuration. Therefore, without loss of generality, we can fix the mutual distances. Let  $d_{jk}=\sqrt{3}$. Then  the potential energy is $V(x)=(m_1m_2+m_1m_3+m_2m_3)/(\sqrt{3})^a$, the moment of inertia is $x^TMx=3(m_1m_2+m_1m_3+m_2m_3)/\mu$, and $\omega_1^2=a\mu/(\sqrt{3})^{a+2}$.
Assume that  the particles are at positions $x_j=x'_j-c$ where $c$ is the center of mass, and
\[
x_1'=(1,0) \quad x_2'=(-1/2,\sqrt{3}/2)\quad x_3'=(-1/2,-\sqrt{3}/2)
\]
Since the formulas for $A$ are translation invariant the $x'_{jk}$ can be used instead of the $x_{jk}$ and one finds that $C=\omega_1^{-2}M^{-1}A$ is:
\[C=\left [\begin{array}{ccc}
C_{11}&C_{12}&C_{13}\\
C_{21}&C_{22}&C_{23}\\
C_{31}&C_{32}&C_{33}\\
\end{array}\right ]\]
where 

{\tiny\[
\begin{split}
&C_{11}=\frac{1}{4\mu}\left [\begin{array}{cc}
(2+3a)(m_2+m_3)&\sqrt{3}(a+2)(m3-m2)\medskip\\
\sqrt{3}(a+2)(m_3-m_2)&-(2-a)(m_2+m_3)
\end{array}\right ], ~\quad
 C_{12}=\frac{1}{4\mu}\left [\begin{array}{cc}
-(2+3a)m_2&\sqrt{3}(a+2)m_2\medskip\\
\sqrt{3}(a+2)m_2&(2-a)m_2
\end{array}\right ]\bigskip\\
&C_{13}=\frac{1}{4\mu}\left [\begin{array}{cc}
-(2+3a)m_3&-\sqrt{3}(a+2)m_3\medskip\\
-\sqrt{3}(a+2)m_3&(2-a)m_3
\end{array}\right ], \quad\qquad\qquad\qquad\quad C_{21}=\frac{1}{4\mu}\left [\begin{array}{cc}
-(2+3a)m_1&\sqrt{3}(a+2)m_1\medskip\\
\sqrt{3}(a+2)m_1&(2-a)m_1
\end{array}\right ]\bigskip\\
&C_{22}=\frac{1}{4\mu}\left [\begin{array}{cc}
(2+3a)m_1-4m_3&-\sqrt{3}(a+2)m_1\medskip\\
-\sqrt{3}(a+2)m_1&(a-2)m_1+4(1+a)m_3
\end{array}\right ], \quad C_{23}=\frac{1}{4\mu}\left [\begin{array}{cc}
4m_3&0\medskip\\
0&-4(1+a)m_3
\end{array}\right ]\bigskip\\
&C_{31}=\frac{1}{4\mu}\left [\begin{array}{cc}
-(2+3a)m_1&-\sqrt{3}(a+2)m_1\medskip\\
-\sqrt{3}(a+2)m_1&(2-a)m_1
\end{array}\right ], \qquad\qquad\qquad\quad\quad C_{32}=\frac{1}{4\mu}\left [\begin{array}{cc}
4m_2&0\medskip\\
0&-4(1+a)m_2
\end{array}\right ]
\end{split}
\]}

and

{\tiny \[
C_{33}=\frac{1}{4\mu}\left [\begin{array}{cc}
(2+3a)m_1-4m_2&\sqrt{3}(a+2)m_1\medskip\\
\sqrt{3}(a+2)m_1&(a-2)m_1+4(1+a)m_2
\end{array}\right].
\]}

%{\tiny\[
%\frac{1}{4\mu} \left [
%\begin{array}{cccccc}
%(2+3a)(m_2+m_3)&\sqrt{3}(a+2)(m3-m2)&-(2+3a)m_2&\sqrt{3}(a+2)m_2&-(2+3a)m_3&-(a+2)\sqrt{3}\medskip\\
%\sqrt{3}(a+2)(m_3-m_2)&-(2-a)(m_2+m_3)&\sqrt{3}(a+2)m_2&(2-a)m_2&-\sqrt{3}(a+2)m_3&(2-a)m_3\medskip\\
%-(2+3a)m_1&\sqrt{3}(a+2)m_1&(2+3a)m_1-4m_3&-\sqrt{3}(a+2)m_1&4m_2&0\medskip\\
%\sqrt{3}(a+2)m_1&(2-a)m_1&-\sqrt{3}(a+2)m_1&(a-2)m_1+4(1+a)m_3&0&-4(1+a)m_2\medskip\\
%-(2+3a)m_1&-(a+2)\sqrt{3}&4m_2&0&(2+3a)m_1-4m_2&\sqrt{3}(a+2)m_1\medskip\\
%-\sqrt{3}(a+2)m_1&(2-a)m_1&0&-4(1+a)m_2&\sqrt{3}(a+2)m_1&(a-2)m_1+4(1+a)m_2\\
%\end{array}\right ]
%\]}

Four  eigenvalues of $C$ are $0,0,-1,a+1$ that we found in  subsection 3.1. Let us denote  with $\eta$ and $\xi$ the remaining eigenvalues. The eigenvalues $\eta$ and $\xi$ can be determined, however we will only need to know the value of  $\eta+\xi$ and $\eta\xi$.
The trace of $C$ is $a+1-1+\eta+\xi=2a$, i.e. $\eta+\xi=a$, thus 
\[\alpha=2-\xi-\eta=2-a.\]   
To find $\beta=(\eta+1)(\xi+1)=\eta\xi+1+a$ we need to compute $\eta\xi$.
%$\beta=(\eta+1)(\xi+1)=\eta\xi+2$  was a typo!!! %

The sum of all possible product of pairs of eigenvalues is $a^2-a-1+\eta\xi$. This sum can also be found as one-half the difference of the square of the trace of the matrix above and the trace of its square. This computation yields
\[ \frac{1}{4\mu^2}[(11a^2-4a-4)(m_1m_2+m_1m_3+m_2m_3)+(4a^2-8a-8)(m_1^2+m_2^2+m_3^2)],\] 
and hence
\[\beta=\frac 3 4 \frac{(a^2+4a+4)(m_1m_2+m_1m_3+m_2m_3)}{\mu^2}.\]
Since $\alpha>0$ and $\beta>0$ the equilateral triangle is spectrally stable if and only if $\alpha^2-4\beta\geq 0$, that is
\beq
\frac{m_1m_2+m_1m_3+m_2m_3}{(m_1+m_2+m_3)^2}\leq\frac{1}{3}\left (\frac{2-a}{2+a}\right)^2.
\label{routhinequality}
\eeq
%The inequality above was first found by Routh. 
Let $f(m_1,m_2,m_3)$ be the left hand side of equation (\ref{routhinequality}). Then the level curves  $f(m_1,m_2,m_3)=c$ can be  plotted on the triangle of normalized masses, see Figure \ref{triangle}. The level curves for levels less than $1/4$ consist of three curves (arcs of circle) near the corners of the triangle, while the levels for  $c>1/4$ determine a single closed curve, a circle centered in the orthocenter of the equilateral triangle. The value $1/4$ corresponds to $a=14-8\sqrt{3}\simeq 0.14359354$.

Figure \ref{triangle}, that depicts the triangle of normalized masses for the homogeneous potential $V$ when $14-8\sqrt{3}<a<2$, can be interpreted in the following way . The shaded region represents the region of stability, i.e. the values of the masses for which the triangle solutions are stable. Thus, when  $14-8\sqrt{3}<a<2$,  the Lagrangian triangle solutions are stable  only for the mass triplets in the corners of the triangle, i.e.  when one mass is considerably larger than the other two. In the Newtonian case, namely when $a=1$, they are unstable for the majority of mass triplets.
 
Figure \ref{triangle1} depicts the triangle of normalized masses when $0<a<14-8\sqrt{3}$. As before the shaded region represents the region of stability. The white disk is the region of instability.
In this case the relative equilibria are  stable not only when there is a dominant mass but also when one of the masses is considerably smaller than the other two. This shows that one cannot extend Moeckel's conjecture to homogeneous potentials.
Moreover as  $a$ decreases the radius of the disk shrinks and the region of stability becomes larger. However if $a>0$ the case where $m_1=m_2=m_3$ is always unstable.

%----------------------------------------------------------
%\begin{center}
%\begin{figure}[t!]
%\begin{minipage}[t]{14cm}
%\parbox[t]{6.5cm}{
%\resizebox{!}{5.5cm}{\includegraphics{fig1.eps}}
%\vskip 0.35 cm \begin{center} {\footnotesize(a)} \end{center}} \hfill
%\parbox[t]{6.5cm}{
%\resizebox{!}{5.5cm}{\includegraphics{fig3.eps}}
%\begin{center} {\footnotesize(b)} \end{center}} \\ \mbox{}
%\end{minipage}
%\vskip -.3 cm
%\caption{(a) Some level curves $f(m_1,m_2,m_3)=c$ in the triangle of normalized masses. (b) fghfghhh}
%\label{triangle}
%\end{figure}
%\end{center}
%---------------------------------------------------------
%\begin{figure}[t]
%\begin{center}
%\resizebox{!}{6cm}{\includegraphics{fig1.eps}}
%\end{center}
%\caption{Some level curves $f(m_1,m_2,m_3)=c$ in the triangle of normalized masses.}
%\label{triangle}
%\end{figure}
%----------------------------------------------------

%%%%%%%%%%%%%%%%%%%%%%%%%%%%%%%%%%%%%%%
\section{Quasihomogeneous Potentials}
%%%%%%%%%%%%%%%%%%%%%%%%%%%%%%%%%%%%%%%
%%%%%%%%%%%%%%%%%%%%%%%%%%
\subsection{General Case}
%%%%%%%%%%%%%%%%%%%%%%%%%%%%%%%%%%%%%%
Consider  the potential $U(x)$. The matrix $\bar A =D\nabla U(x)$ is of the form
\[\bar A=\left [\begin{array}{ccc}
\bar A_{11}& \cdots& \bar A_{1n}\\
\vdots   &  &\vdots\\
\bar A_{n1}&\cdots&\bar A_{nn}\\
\end{array}\right ]\]
where $\bar A_{ij}$ is the $2\times 2$ matrix given by
\beq
\begin{split} 
\bar A_{jk}&=A_{jk}+B_{jk}, \quad \mbox{if}\quad j\neq k\\
\bar A_{kk}&=-\sum_{j\neq k}\bar A_{jk}
\end{split}
\eeq
where
\beq
A_{jk}=a\frac{m_im_k}{d_{jk}^{a+2}} \left [I-(a+2) u_{jk} u_{jk}^T \right], \quad j\neq k
\eeq
and 
\beq
B_{jk}=b\frac{m_im_k}{d_{jk}^{b+2}} \left [I-(b+2) u_{jk} u_{jk}^T \right]. \quad j\neq k
\eeq
As in the case of the homogeneous potential one can apply Proposition \ref{prop} taking the vectors
$v=(1,0,\cdots,1,0)$, $w=(0,1,\cdots,0,1)$ that are in the kernel of $\bar A$. Again one finds the eigenvalues $\mu=\pm i$

Now consider $v=x$ (i.e. a relative equilibrium for $V$, $W$ and $U$)  and $w=Jx$. Equation (\ref{hom}) with the homogeneity of the potentials $V$ and $W$ gives:
\beq
\begin{split}
 M^{-1}D\nabla V(x)v &=-(a+1)M^{-1}\nabla V(x)\\&=(a+1)M^{-1}\omega_1^2Mx=(a+1)\omega_1^2v,
\end{split}
\eeq
and 
\beq
\begin{split}
M^{-1}D\nabla W(x)v &=-(b+1)M^{-1}\nabla W(x)\\&=(b+1)M^{-1}\omega_2^2Mx=(b+1)\omega_2^2v.
\end{split}
\eeq
Consequently
\beq
\ M^{-1}D\nabla U(x)v=M^{-1}[D\nabla V(x)+D\nabla W(x)]v=[(a+1)\omega_1^2+(b+1)\omega_2^2]v
\eeq
so $v$ is an eigenvector of $M^{-1}\bar A$ with eigenvalue $[(a+1)\omega_1^2+(b+1)\omega_2^2]$.
On the other hand,  $w=(R^{-1}(0))'x$, as in the case of the homogeneous potential. 
Now
\[
\nabla U(R^{-1}(\theta)x)=R^{-1}(\theta)\nabla U(x)
\]
because of the rotation invariance of the potential. Differentiating at $\theta=0$ gives:
\beq
M^{-1}D\nabla U(x)w=M^{-1}J\nabla U(x)=-M^{-1}J\omega^2Mx=-\omega^2w
\eeq
where $\omega^2=\omega_1^2+\omega_2^2$. Therefore $w$ is an eigenvector with eigenvalue $-\omega^2$. Choosing  
\beq\eta=[(a+1)\omega_1^2+(b+1)\omega_2^2]/\omega^2=1+ \frac{a\omega_1^2+b\omega_2^2}{\omega^2}\label{eta}\eeq and $\xi=-1$ we find that 
\beq\alpha=2-\frac{a\omega_1^2+b\omega_2^2}{\omega^2}\eeq
and the roots of $Q(z)$ are positive if $[a\omega_1^2+b\omega_2^2]/\omega^2>2$.
Thus we  have proved the following
\begin{theorem}
Any  simultaneous relative equilibrium of $V(x)$, $W(x)$ and $U(x)$  is spectrally unstable if $[a\omega_1^2+b\omega_2^2]/\omega^2>2$.
\label{thquasihom}
\end{theorem}

%%%%%%%%%%%%%%%%%%%%%%%%%%
\subsection{Lagrangian Triangle Relative Equilibria}
%%%%%%%%%%%%%%%%%%%%%%%%%%%%
For $n=3$ the equilateral triangle configurations with the center of mass at the origin are simultaneous  relative equilibria for the potentials  $V(x)$, $W(x)$ and $U(x)$. They are relative equilibria for any choice of the masses $m_1$, $m_2$ and $m_3$. 

In this case we assume that  the particles are at positions $x_j=x'_j-c$ where $c$ is the center of mass, and
\[
x_1'=(r,0) \quad x_2'=(-r/2,r\sqrt{3}/2)\quad x_3'=(-r/2,-r\sqrt{3}/2)
\]
where now we cannot fix the size of the configuration since the potential is not homogeneous.
Consequently the  mutual distances are $d_{jk}=r_0$, where $r_0=\sqrt{3}r$, the potential energy is $U(x)=(m_1m_2+m_1m_3+m_2m_3)(1/r_0^a+1/r_0^b)$ and  the moment of inertia is $x^TMx=r_0^2(m_1m_2+m_1m_3+m_2m_3)/\mu$. Moreover  $\omega_1^2=a\mu/(r_0)^{a+2}$, $\omega_2^2=b\mu/(r_0)^{b+2}$ and $\omega^2=\omega_1^2+\omega_2^2$. As in the homogeneous case the   $x'_{jk}$ can be used to compute  $\bar C=\omega^{-2}M^{-1}\bar A$, that is of the form

\[\bar C=\left [\begin{array}{ccc}
\bar C_{11}&\bar C_{12}&\bar C_{13}\\
\bar C_{21}&\bar C_{22}&\bar C_{23}\\
\bar C_{31}&\bar C_{32}&\bar C_{33}\\
\end{array}\right ]\]
where
{\tiny
\[\bar C_{11}=\frac{1}{4\omega^2}\left [\begin{array}{cc}
\frac{a(2+3a)(m_2+m_3)}{r_0^{a+2}}+\frac{b(2+3b)(m_2+m_3)}{r_0^{b+2}}&\frac{a\sqrt{3}(a+2)(m3-m2)}{r_0^{a+2}}+\frac{b\sqrt{3}(b+2)(m3-m2)}{r_0^{b+2}}\medskip\\
\frac{a\sqrt{3}(a+2)(m_3-m_2)}{r_0^{a+2}}+\frac{b\sqrt{3}(b+2)(m_3-m_2)}{r_0^{b+2}}&-\frac{a(2-a)(m_2+m_3)}{r_0^{a+2}}-\frac{b(2-b)(m_2+m_3)}{r_0^{b+2}}
\end{array}\right ]\]

\[\bar C_{12}=\frac{1}{4\omega^2}\left [\begin{array}{cc}
-\frac{a(2+3a)m_2}{r_0^{a+2}}-\frac{b(2+3b)m_2}{r_0^{b+2}}&\frac{a \sqrt{3}(a+2)m_2}{r_0^{a+2}}+\frac{b \sqrt{3}(b+2)m_2}{r_0^{b+2}}\medskip\\
\frac{a\sqrt{3}(a+2)m_2}{r_0^{a+2}}+\frac{b\sqrt{3}(b+2)m_2}{r_0^{b+2}}&\frac{a(2-a)m_2}{r_0^{a+2}}+\frac{b(2-b)m_2}{r_0^{b+2}}\end{array}\right ]
\]

\[\bar C_{13}=\frac{1}{4\omega^2}\left [\begin{array}{cc}
-\frac{a(2+3a)m_3}{r_0^{a+2}}-\frac{b(2+3b)m_3}{r_0^{b+2}}&-\frac{a\sqrt{3}(a+2)m_3}{r_0^{a+2}}-\frac{b\sqrt{3}(b+2)m_3}{r_0^{b+2}}\medskip\\
-\frac{a\sqrt{3}(a+2)m_3}{r_0^{a+2}}-\frac{b\sqrt{3}(b+2)m_3}{r_0^{b+2}} & \frac{a(2-a)m_3}{r_0^{a+2}}+\frac{b(2-b)m_3}{r_0^{b+2}}
\end{array}\right ]
\]

\[\bar C_{21}=\frac{1}{4\omega^2}\left [\begin{array}{cc}
-\frac{a(2+3a)m_1}{r_0^{a+2}}-\frac{b(2+3b)m_1}{r_0^{b+2}}&\frac{a \sqrt{3}(a+2)m_1}{r_0^{a+2}}+\frac{b \sqrt{3}(b+2)m_1}{r_0^{b+2}}\medskip\\
\frac{a\sqrt{3}(a+2)m_1}{r_0^{a+2}}+\frac{b\sqrt{3}(b+2)m_1}{r_0^{b+2}}&\frac{a(2-a)m_1}{r_0^{a+2}}+\frac{b(2-b)m_1}{r_0^{b+2}}\end{array}\right ]
\]

\[
\bar C_{22}=\frac{1}{4\omega^2}\left [\begin{array}{cc}
\frac{a((2+3a)m_1-4m_3)}{r_0^{a+2}}+\frac{b((2+3b)m_1-4m_3)}{r_0^{b+2}}&-\frac{a\sqrt{3}(a+2)m_1}{r_0^{a+2}}-\frac{b\sqrt{3}(b+2)m_1}{r_0^{b+2}}\medskip\\
-\frac{a\sqrt{3}(a+2)m_1}{r_0^{a+2}}-\frac{b\sqrt{3}(b+2)m_1}{r_0^{b+2}}&\frac{(a(a-2)m_1+4(1+a)m_3)}{r_0^{a+2}}+\frac{(b(b-2)m_1+4(1+b)m_3)}{r_0^{b+2}}
\end{array}\right ]\]

\[
\bar C_{23}=\frac{1}{4\omega^2}\left [\begin{array}{cc}
\frac{4am_3}{r_0^{a+2}}+\frac{4bm_3}{r_0^{b+2}}&0\medskip\\
0&-\frac{4a(1+a)m_3}{r_0^{a+2}}-\frac{4b(1+b)m_3}{r_0^{b+2}}
\end{array}\right ]\]

\[\bar
C_{31}=\frac{1}{4\omega^2}\left [\begin{array}{cc}
-\frac{a(2+3a)m_1}{r_0^{a+2}}-\frac{b(2+3b)m_1}{r_0^{b+2}}&-\frac{a\sqrt{3}(a+2)m_1}{r_0^{a+2}}-\frac{b\sqrt{3}(b+2)m_1}{r_0^{b+2}}\medskip\\
-\frac{a\sqrt{3}(a+2)m_1}{r_0^{a+2}}-\frac{b\sqrt{3}(b+2)m_1}{r_0^{b+2}} & \frac{a(2-a)m_1}{r_0^{a+2}}+\frac{b(2-b)m_1}{r_0^{b+2}}
\end{array}\right ]
\]

\[\bar C_{32}=\frac{1}{4\omega^2}\left [\begin{array}{cc}
\frac{4am_2}{r_0^{a+2}}+\frac{4bm_2}{r_0^{b+2}}&0\medskip\\
0&-\frac{4a(1+a)m_2}{r_0^{a+2}}-\frac{4b(1+b)m_2}{r_0^{b+2}}
\end{array}\right ]\]

\[\bar C_{33}=\frac{1}{4\omega^2}\left [\begin{array}{cc}
\frac{a((2+3a)m_1-4m_2)}{r_0^{a+2}}+\frac{b((2+3b)m_1-4m_2)}{r_0^{b+2}}&\frac{a\sqrt{3}(a+2)m_1}{r_0^{a+2}}\frac{b\sqrt{3}(b+2)m_1}{r_0^{b+2}}\medskip\\
\frac{a\sqrt{3}(a+2)m_1}{r_0^{a+2}}+\frac{b\sqrt{3}(b+2)m_1}{r_0^{b+2}}& \frac{a((a-2)m_1+4(1+a)m_2)}{r_0^{a+2}}+\frac{b((b-2)m_1+4(1+b)m_2)}{r_0^{b+2}}
\end{array}\right].
\]}
Four  eigenvalues of $\bar C$ were found in subsection 4.1: $0,0,-1$ and $1+\frac{a\omega_1^2+b\omega_2^2}{\omega^2}$. Or, using the values of $\omega_1,\omega_2$ and $\omega$ found in the case of three bodies $0,0,-1$ and 
\[1+\frac{a^2r_0^{b+2}+b^2r_0^{a+2}}{ar_0^{b+2}+br_0^{a+2}}.\]
Theorem \ref{thquasihom} applied to the triangle solutions shows that they are  spectrally unstable for any value of the masses $m_1,m_2$ and $m_3$ when
\beq
\frac{(a^2-2a)r_0^{b+2}+(b^2-2b)r_0^{a+2}}{ar_0^{b+2}+br_0^{a+2}}>0.
\label{firstcondition}
\eeq
Let us denote the  remaining eigenvalues of $\bar C$ with $\bar \eta$ and $\bar \xi$. We will only need to know the value of $\bar \eta+\bar \xi$ and $\bar\eta\bar\xi$. 
Taking the trace of $\bar C$ gives
\[
\mbox{Tr}(\bar C)=\frac{a^2r_0^{b+2}+b^2r_0^{a+2}}{ar_0^{b+2}+br_0^{a+2}}+\bar\eta+\bar\xi=2\frac{a^2r_0^{b+2}+b^2r_0^{a+2}}{ar_0^{b+2}+br_0^{a+2}}
\]
or 
\[
\bar\eta+\bar\xi=\frac{a^2r_0^{b+2}+b^2r_0^{a+2}}{ar_0^{b+2}+br_0^{a+2}}.
\]
In the following  we will use $\delta$ to denote the right hand side of the previous equation. The first parameter in Proposition \ref{prop} is
\[
\alpha=2-\bar\eta-\bar\xi=-\frac{(a^2-2a)r_0^{b+2}+(b^2-2b)r_0^{a+2}}{ar_0^{b+2}+br_0^{a+2}}.
\]
To find $\beta=(\bar \eta+1)(\bar \xi +1)=1+\delta+\bar\eta\bar\xi$ one needs to compute $\bar\eta\bar\xi$. In order to do that one can observe that the sum of all possible pairs  of eigenvalues of $\bar C$ can be found in two ways. Direct computations give $-1-\delta+\delta^2+\bar\eta\bar\xi$. On the other hand one can find this sum as $T=(1/2)[\mbox{Tr}(\bar C)^2-\mbox{Tr}((\bar C)^2)]$ where 
\beq
\begin{split}
T=&\left [(m_1^2+m_2^2+m_3^2)(r_0^{2a+4}(4b^4-8b^3-8b^2)+r_0^{2b+4}(4a^4-8a^3-8a^2)) \right.\\
&+(m_1^2+m_2^2+m_3^2)(r_0^{a+b+4}(8a^2b^2-8ab^2-8a^2b-16ab)))\\
&+(m_1m_2+m_2m_3+m_1m_3)\\
&\times(r_0^{2a+4}(11b^4-4b^3-4b^2)+r_0^{2b+4}(11a^4-4a^3-4a^2))\\
& \left. +(m_1m_2+m_2m_3+m_1m_3)(r_0^{a+b+4}(22a^2b^2-4ab^2-4a^2b-8ab))\right ]\\
&\times(4\mu^2(ar_0^{b+2}+ar_0^{a+2})^2)^{-1}
\end{split}
\eeq
which leads to $\beta=2(1+\delta)-\delta^2+T$, or explicitly, after some simplifications to
\beq
\beta=\frac{3(m_1m_2+m_1m_3+m_2m_3)}{4\mu^2}\left(\frac{b(b+2)r_0^{a+2}+a(a+2)r_0^{b+2}}{br_0^{a+2}+ar_0^{b+2}}\right)^2.
\eeq
%\beq
%\frac{m_1m_2+m_1m_3+m_2m_3}{(m_1+m_2+m_3)^2}<\frac{1}{3}\left (\frac{b(b-2)r_0^{a+2}+a(a-2)r_0^{b+2}}{b(b+2)r_0^{a+2}+a(a+2)r_0^{b+2}}\right)^2%\label{generalrouthinequality}
%\eeq
Since $\beta>0$ the equilateral triangle relative equilibrium is spectrally stable if and only if 
$\alpha>0$ and
\beq
\frac{m_1m_2+m_1m_3+m_2m_3}{(m_1+m_2+m_3)^2}\leq\frac{1}{3}\left (\frac{b(b-2)r_0^{a-b}+a(a-2)}{b(b+2)r_0^{a-b}+a(a+2)}\right)^2.
\label{generalrouthinequality}
\eeq
This generalizes Routh stability condition to the case of quasihomogeneous potentials.

We now want to show that, in this case, the stability depends, for certain values of the masses, on the size of the configuration, i.e. of the equilateral triangle. In order to do that let $f(m_1,m_2,m_3)$ be the left hand side and $g(a,b,r_0)$ be the right hand side of equation (\ref{generalrouthinequality}) and consider   the limits of $g(a,b,r_0)$,  as $r_0\rightarrow 0$ and $r_0\rightarrow \infty$.
The first limit is
\[\lim_{r_0 \to 0}g(a,b,r_0)=\frac{1}{3}\left (\frac{a-2}{a+2}\right)^2\]
while the second is
\[\lim_{r_0 \to \infty}g(a,b,r_0)=\frac{1}{3}\left (\frac{b-2}{b+2}\right)^2.\]
With the preparations above we are well on our way to proving the following

\begin{theorem}
Consider the  Lagrange triangle solution for  the quasihomogeneous potential $U$.
\begin{enumerate}

\item If $0<b<a<2$ it is
\begin{enumerate}
\item Unstable when $f>1/3((b-2)/(b+2))^2$
\item Spectrally stable when $f\leq 1/3((a-2)/(a+2))^2$
\item Unstable when $1/3((a-2)/(a+2))^2<f\leq 1/3((b-2)/(b+2))^2$ and $r<r_0^*$, spectrally stable when $1/3((a-2)/(a+2))^2<f<1/3((b-2)/(b+2))^2$ and $r\geq r_0^*$ for some $r_0^*\in \mathbb{R}^+$ (where  $r_0^*=r_0^*(f)$).
\end{enumerate}
\item If  $0<b<2<a$ it is
\begin{enumerate}
\item Unstable when $f>(1/3)[(b-2)/(b+2)]^2$
\item Unstable when $0<f\leq(1/3)[(b-2)/(b+2)]^2$ and  $r_0<z_1^*$ for some $z_1^*\in \mathbb{R}^+$ (where  $z_1^*=z_1^*(f)$).
\item Spectrally stable  when $0<f\leq(1/3)[(b-2)/(b+2)]^2$ and $r_0\geq z_1^*$.
\end{enumerate}
\item if $2<b<a$ it is unstable for any value of the masses.
\end{enumerate}
\end{theorem}

%---------------------------------------------------------
%\begin{figure}[t]
%\begin{center}
%\resizebox{!}{6cm}{\includegraphics{fig2.eps}}
%\end{center}
%\caption{Region of stability in the triangle of normalized masses for the quasihomogeneous potential $U$ with $0<b<a<14-8\sqrt{3}$. The equilateral triangle is stable for the mass triplets in $D_1$ and spectrally stable for the ones in $D_3$. When the mass triplet is in  $D_2$ the stability depends on the size of the triangle: it is unstable for $r<r_0^*$ and stable for $r>r_0^*$. }
%\label{triangle2}
%\end{figure}
%----------------------------------------------------
%-------------------------------------------------------------------------------------
\begin{figure}[t]
  \begin{center}
    \mbox{
      \subfigure[]{\resizebox{!}{5.5cm}{\includegraphics{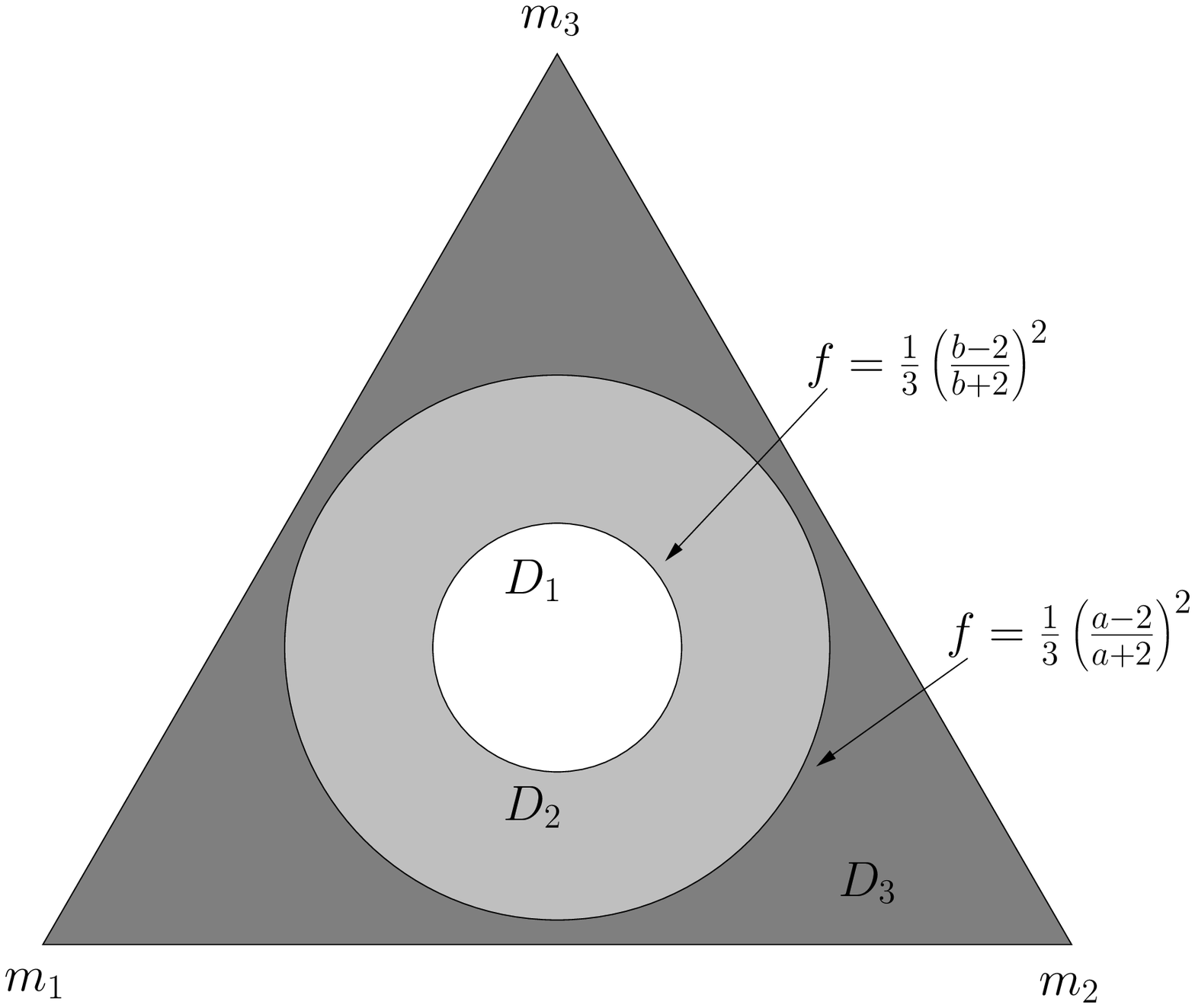}} \label{triangle2}} \quad
      \subfigure[]{\resizebox{!}{5.5cm}{\includegraphics{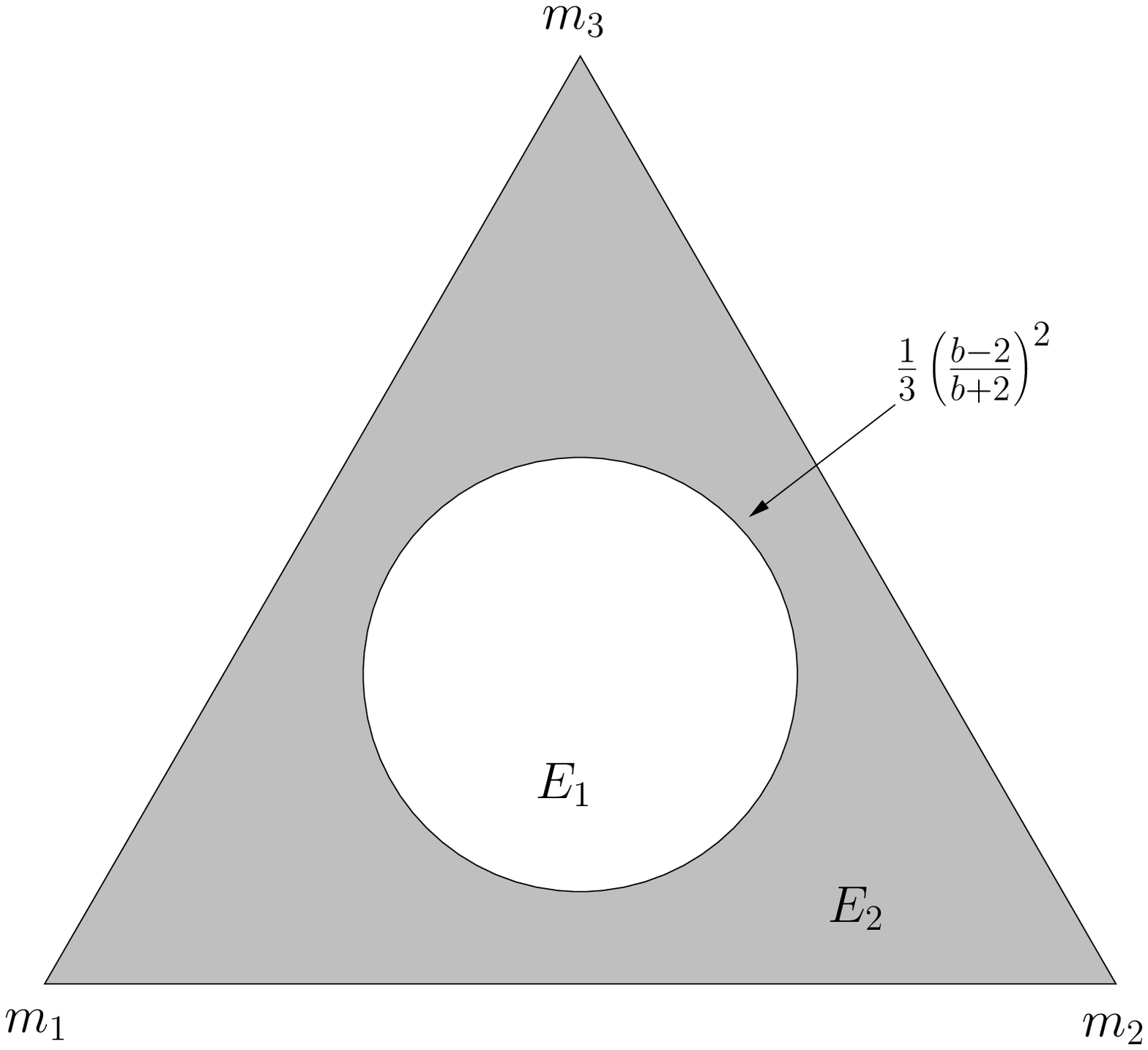}} \label{triangle3}}} 
    \caption{(a) Regions of stability in the triangle of normalized masses for the  potential $U$ with $0<b<a<14-8\sqrt{3}$. The equilateral triangle is unstable for the mass triplets in $D_1$ and spectrally stable for the ones in $D_3$. When the mass triplet is in  $D_2$ the stability depends on the size of the triangle: it is unstable for $r<r_0^*$ and stable for $r\geq r_0^*$. (b) Regions of stability in the triangle of normalized masses for the  potential $U$ when $0<b<14-8\sqrt{3}$, $a>2$. The equilateral triangle is unstable for the mass triplets in $E_1$. When the mass triplet is in  $E_2$ the stability depends on the size of the triangle: it is unstable for $r<z_1^*$ and stable for $r\geq z_1^*$}
  \end{center}
\end{figure}
%-----------------------------------------------------------------------------------------------

\begin{proof}
First we prove part (a) of the theorem. When $0<b<a<2$, $\alpha>0$ then  $\lim_{r_0 \to 0}g(a,b,r_0)<\lim_{r_0 \to \infty}g(a,b,r_0)$, since $1/3((a-2)/(a+2))^2$ is a monotonically decreasing function for $0<a<2$.
Therefore from inequality (\ref{generalrouthinequality}) follows that when the masses satisfy 
\[
\frac{1}{3}\left (\frac{a-2}{a+2}\right)^2<\frac{m_1m_2+m_1m_3+m_2m_3}{(m_1+m_2+m_3)^2}\leq \frac{1}{3}\left (\frac{b-2}{b+2}\right)^2
\]
the relative equilibrium is spectrally unstable as $r_0\rightarrow 0$ but spectrally stable when $r_0\rightarrow \infty$. 
Moreover
\begin{eqnarray}
\begin{split}
\frac{\partial g}{\partial r_0}=&\frac 2 3 \frac{b(a-b)r_0^{a-b}[(b^2-2b)r^{a-b}+(a^2-2a)]}{b(b+2)r_0^{a-b}+a(a+2)}\\
&\times\left[(b-2)+(b+2)\frac{(b^2-2b)r^{a-b}+(a^2-2a)}{b(b+2)r_0^{a-b}+a(a+2)} \right]
\label{diffg}
\end{split}
\end{eqnarray}
is always negative if $a<2$, $b<2$ and $a<b$. Therefore $g(r_0)$ is  a monotonically decreasing function of $r_0$. Consequently there exists an unique  $r_0^*$ such that the triangle is unstable for $r<r_0^*$ and spectrally stable for $r\geq r_0^*$. In other words the relative equilibrium is unstable  when the equilateral triangle is small but it is spectrally stable if the triangle is large. 

On the other hand, since $g(r_0)$ is a monotonically decreasing function of $r_0$ then $1/3((a-2)/(a+2))^2<g<1/3((b-2)/(b+2))^2$. Thus  the triangle is spectrally stable when  $f\leq1/3((a-2)/(a+2))^2$ and it is unstable when $f>1/3((b-2)/(b+2))^2$. Figure \ref{triangle2} depicts the regions of stability and instability for $0<b<a<14-8\sqrt{3}$.

We now prove part (b) of the theorem. Observe that $\alpha>0$ in this case gives $r_0>z^*$ where 
\[z^*=\left (-\frac{a^2-2a}{b^2-2b} \right )^{\frac{1}{a-b}}.\]
Thus the triangle is unstable for any value of the masses when $r_0<z^*$.

On the other  hand $g(a,b,z^*)=0$. Moreover  from equation (\ref{diffg}) one can see that $\partial g/\partial r_0$ is negative  for $r_0<z^*$ and positive for $r_0>z^*$. This shows that $g(a,b,r_0)$ is a monotonically decreasing function of $r_0$ in  the interval $(0,z^*)$ and a monotonically increasing one in $(z^*,\infty)$. Therefore if we fix the values of the masses so that $0<f<(1/3)[(b-2)/(b+2)]^2$ the equation $f=g(a,b,r_0)$ has a unique solution $z_1^*$ in the interval $(z^*,\infty)$. Consequently
$f>g(a,b,r_0)$ for $r_0<z_1^*$ and the triangle is unstable, while  $f\leq g(a,b,r_0)$ for $r_0\geq z_1^*$ and the triangle is spectrally stable. Figure \ref{triangle3} depicts the regions of stability and instability for $0<b<14-8\sqrt{3}$, $a>2$.

The proof of part (c) follows immediately from inequality (\ref{firstcondition}).
\end{proof}

\section*{Acknowledgements} 
I would like to thank Alain Albouy and Rick Moeckel for their advice  and suggestions regarding this work.
%%%%%%%%%%%%%%%%%%%%%%%%%%%%%%%%%
%\section*{References}

\end{document}